\documentclass[twoside,fleqn]{article}
\usepackage{espcrc2,psfig,epsfig}

\newcommand{\be}{\begin{equation}}
\newcommand{\ee}{\end{equation}}

\newcommand{\AmS}{{\protect\the\textfont2
  A\kern-.1667em\lower.5ex\hbox{M}\kern-.125emS}}

\title{Ultra High Energy Cosmic Rays from Cosmological Relics}

\author{V. Berezinsky\address{INFN, Laboratori Nazionali del Gran Sasso,
             I--67010 Assergi (AQ), Italy\\ and Institute for Nuclear 
Research, Moscow, Russia}
}       
\begin{document}

\begin{abstract}
The current status of origin of Ultra High Energy Cosmic Rays (UHECR)
is reviewed, with 
emphasis given to elementary particle solutions to UHECR problem,
namely to Topological Defects and Super-Heavy Dark Matter (SHDM) 
particles.   
The relic superheavy particles are very efficiently produced at
inflation. Being protected by gauge discrete symmetries, they can be
long lived. They are clustering in the Galactic halo, producing thus
UHECR without Greisen-Zatsepin-Kuzmin cutoff.  
Topological Defects can naturally produce particles with energies as 
observed and much higher, but in most cases fail to produce the observed 
fluxes. Cosmic necklaces, monopoles connected by strings and vortons 
are identified as  most plausible sources. The latter two of them are 
also clustering in the halo and their observational predictions are
identical to those of SHDM particles.

\end{abstract}

\maketitle

\section{\bf Introduction }
Ultra High Energy Cosmic Rays (UHECR) is a puzzle of modern physics. 
Its solution needs the new ideas in astrophysics or in elementary 
particle physics. 
\vspace{-3mm}
\begin{figure}[h]
\begin{center}
\psfig{bbllx=60pt, bblly=160pt, bburx=530pt, bbury=616pt,
file=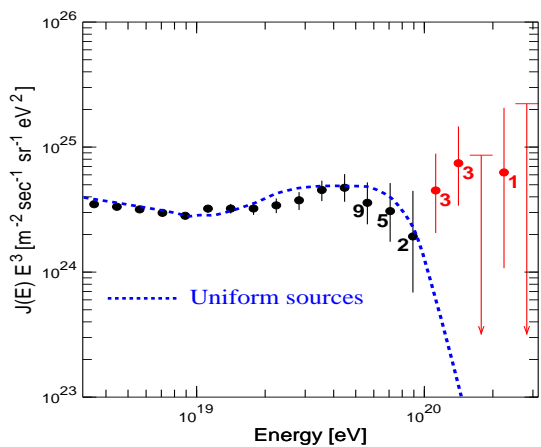, height=6cm, width=7.5cm, clip}
\end{center}
\vspace{-12mm}
\caption{\em  AGASA spectrum compared with ``astrophysical''
spectrum calculated under assumptions that the sources distributed 
uniformly in the Universe and have generation spectrum $\sim E^{-2.3}$}        
\cite{AGASA}
\end{figure}
\vspace{-3mm}

The problem of UHECR is known for more than 30 years. It
consits in observation of primary particles with energies up to 
$2 - 3\cdot 10^{20}~eV$ \cite{obs}. If these particles are
extragalactic protons and their sources are distributed uniformly in
the Universe, their spectrum must expose steepening, which starts at 
energy $E_{bb} \approx 3\cdot 10^{19}~eV$ due to interaction with
microwave photons. This steepening is known as the
Greisen-Zatsepin-Kuzmin (GZK) cutoff \cite{GZK}. The GZK cutoff is not
seen in the observed spectrum. The spectrum of UHECR according to
AGASA observations \cite{AGASA} is shown in Fig.1 together with 
the spectrum calculated for uniform distribution of the sources in 
the Universe under assumption that generation spectrum is 
proportional to $E^{-2.3}$. The excess of the observed events above
the GZK cutoff is clearly seen.
The observational data for UHECR ($E \leq 1\cdot 10^{19}~eV$) 
can be summarized as follows.
\begin{itemize}
\item At $E \geq 10^{19}~eV$ the spectrum is flatter than at
lower energies
and it extends up to $2 - 3\cdot 10^{20}~eV$ (maximum observed
energies).
\item Chemical composition is favoured by protons, though UHE
photons are not excluded as primaries.
\item Data are consistent with isotropy, but 
close angular pairs (doublets) and triplets compose about $20\%$ of
all events at $E \geq 4\cdot 10^{19}~eV$ (22 events in doublets and 
triplets from 92 total \cite{doubl}.  
\end{itemize}
\vspace{1mm}
{\em Galactic origin} of UHECR due to acceleration by sources located in 
the Galactic disc is excluded. Numerical simulations of propagation of 
UHECR in magnetic fields of disc and halo of the Galaxy predict the 
strong anisotropy for particles with rigidity 
$E/Z > 1\cdot 10^{19}~eV$ (\cite{BBDGP} - \cite{HMR}).\\*[1mm] 
{\em Extragalactic protons}, if their sources are distributed uniformly
in the universe, should have GZK cutoff due to pion production on 
microwave radiation (see Fig.1).\\*[1mm]
{\em Extragalactic nuclei} exhibit the cutoff at the same energy 
$E \sim 3\cdot 10^{19}~eV$, mainly due to $e^+e^-$-pair production on 
microwave radiation \cite{BGZ},\cite{BBDGP},\cite{Roul}. \\*[1mm]
{\em Nearby sources} must form a compact group with large overdensity 
of the sources to avoid GZK cutoff \cite{BBDGP}. Local Supercluster (LS) with 
the typical size $R_{LS} \sim 10~Mpc$ is a natural candidate for such
group. The calculations (see \cite{BBDGP}) show that for absence of 
GZK cutoff the LS overdensity $\delta_{LS}>10$ is needed, while the
observed one is $\delta_{LS} \approx 1.4$ \cite{Pee}. Note, that
diffusion propagation due to magnetic field cannot help in softening 
of GZK cutoff.\\*[1mm]
{\em Nearby single source} can provide the absence of GZK cutoff. 
The idea is that powerful sources of UHECR in the Universe are very
rare and by chance we live nearby one of them. Such case 
has been numerically studied  for the burst generation 
of UHECR and their non-stationary diffuse propagation \cite{BDG}. 
Anisotropy can be small even at energies exceeding $1\cdot 10^{20}~eV$.
The calculated cutoff at $E_c \approx 1\cdot 10^{20}~eV$ is 
questioned by existence of two events with energy $2\cdot 10^{20}~eV$.

An interesting case of single source UHECR origin was 
recently proposed in \cite{BiSta}. The physical essence of this model can be
explained in the following way. A nearby single source is the powerful 
radio galaxy M87 in Virgo cluster. UHE particles from it falls to 
gigantic magnetic halo of our Galaxy (with height about $1.5~Mpc$),
where the azimuthal magnetic field diminishes as $1/r$. 
Magnetic field in the halo focuses the highest energy particles to the 
Sun in such way, that arriving particles have isotropical distribution.
Numerical simulations of the trajectories in the
magnetic field, similar to that in the  galactic wind, confirm this
model. This interesting proposal should be further studied taking into
account such phenomena as diffuse radio, X-ray and gamma radiation
produced by high energy electrons diffusing from the Galactic disc.
The calculations of these processes limit the size of
magnetic halo by $3 - 5~kpc$ \cite{Str}.\\*[1mm]   
{\em Acceleration of UHECR} is a problem for astrophysical 
scenarios. Shock acceleration  and unipolar induction are the "standard" 
acceleration mechanisms to UHE, considered in the literature (see
\cite{BBDGP} for a review).
A comprehensive list of possible sources with shock acceleration was
thoroughly studied in ref.(\cite{NMA}) with a conclusion, that maximum energy 
of acceleration does not exceed $10^{19} - 10^{20}~eV$ (see also 
ref.(\cite{Bland}) with a similar conclusion). The most promising 
source from this list is a hot spot in radiogalaxy produced by a jet 
\cite{BiSt,IA,RB}, where maximum energy can reach $\sim 10^{20}~eV$. 
Radiogalaxy M87, considered in \cite{BiSta}, belongs to this
class of sources.\\*[1mm]
{\em Gamma Ray Bursts} (GRB) models offer two new mechanisms of
acceleration to UHE. The first one \cite{Vie} is acceleration by
ultrarelativistic shock. A reflected particle gains at one reflection
the energy $E \sim \Gamma_{sh}^2 E_i$, where 
$\Gamma_{sh} \sim 10^2 - 10^3$ is the Lorentz factor of the shock and 
$E_i$ is initial energy of a particle. The second cycle of such
acceleration has extremely low probability to occur \cite{GaAch,Ost} 
and therefore 
to produce the particles with $E \sim 10^{20}~ eV$, this mechanism
must operate in the space filled by pre-accelerated particles with 
energies $E_i> 10^{14}~eV$. 

The second mechanism \cite{Wax} works in the model with multiple shocks. 
The collisions of the shocks produces the turbulence where the
particles are accelerated by Fermi II mechanism. The turbulent velocities
are mildly relativistic in the fireball rest system. The maximum energy 
in the rest system , $E_{max}' \sim e H_0'l_0'$, is boosted by Lorentz 
factor $\Gamma$ of fireball in laboratory system (here $l_0'$ and $H_0'$ 
are the maximum linear scale of turbulence with coherent magnetic
field $H_0'$ there).
Taking for $H_0'$ equipartition value, one obtains $E_{max} \sim 10^{20}~eV$
in the laboratory system. This mechanism faces two
problems. Actually the maximum energy is somewhat less than 
$1\sim 10^{20}~eV$ \cite{RaMe}, if  acceleration time is evaluated 
more realistically. It diminishes the energy of GZK cutoff in
the diffuse spectrum, because it is formed by the particles with 
production energies higher than the observed ones. The most serious
problem, however, is that the produced
flux of accelerated particles suffer the adiabatic energy losses \cite{RaMe}.
\\*[1mm]
{\em In summary}, the acceleration (astrophysical) scenarios are somewhat
disfavoured, but not excluded. Apart from them, many elementary
particle solutions were proposed to solve UHECR puzzle. Among them
there is such an extreme proposal as breaking the Lorentz invariance
\cite{LI}, light gluino as the lightest supersymmetric particle and 
UHE carrier \cite{gluino}, UHE neutrinos producing UHECR due to
resonance interaction with the dark matter neutrinos \cite{neutrino} 
and some other suggestions. In this paper I will review two most
conservative sources of UHECR of non-accelerator origin: Superheavy 
Dark Matter (SHDM) and Topological Defects (TD).

\section{UHECR from Superheavy Dark Matter}

Superheavy Dark Matter (SHDM) as a source of UHECR was first
suggested in refs.(\cite{BKV,KR}). 
SHDM particles with masses larger than $10^{13}~GeV$ are accumulated
in the Galactic halo \cite{BKV} with overdensity $\sim 10^5$ and hence
UHECR produced in their decays do not exhibit the GZK 
cutoff. The other observational signatures of this model are dominance 
of UHE photons \cite{BKV} and anisotropy connected with non-central position of
the Sun in the Galactic halo \cite{DuTi,BBV}.\\*[2mm]
\noindent
{\em Production of SHDM}\\
SHDM particles are very efficiently produced by the various mechanisms at 
post-inflationary epochs. This common feature has a natural  
explanation. The SHDM particles due to their tremendous mass had never
been in the
thermal equilibrium in the Universe and never were relativistic. Their
mass density diminished as $\sim 1/a^3$, while for all other particles 
it diminishes much faster as $\sim 1/a^4$, where $a$ is the scaling 
factor of the Universe. When normalized at inflationary epoch,
$a_i=1$, $\;\; a(t)$ reaches enormous value at large $t$. It is
enough to produce negligible amount of superheavy particles in the
post-inflationary epoch in order to provide $\Omega_X \sim 1$ now.
Actually, in most cases one meets a problem of 
overproduction of SHDM particles (further on we shall refer to  them 
as to X-particles).
 
One very general mechanism of X-particle production is given by
creation of particles in time-variable classical field. In our case
it can be inflaton field $\phi$ or gravitational field. In case of
inflaton field the direct coupling of X-particle (or some intermediate  
particle $\chi$) with inflaton is needed,{\em e.g.} $g^2\phi^2 X^2$ or 
$g^2\phi^2\chi^2$. The 
intermediate particle $\chi$ then decays to X-particle. In case of 
time-variable gravitational field no coupling of X to inflaton or 
any other particles is needed: X-particles are produced due to their masses.
For the review of above-mentioned mechanisms and references see \cite{KuTkr}.

Super-heavy particles are very efficiently produced at {\em preheating}
\cite{KLS}. This stage, predecessor of {\em reheating}, is caused by 
oscillation of inflaton field after inflation near the minimum of 
the potential. Such oscillating field can non-perturbatively 
(in the regime of broad parametric resonance) produce the
intermediate bosons $\chi$, which  then decay to X-particles. The mass of
X-particles can be one-two orders of magnitude larger than inflaton
mass $m_{\phi}$, which should be about $10^{13}~GeV$ to provide the
amplitude of density fluctuations observed by COBE. 

Another mechanism, more efficient than parametric resonance and 
operating in its absence, is so-called 
{\em instant preheating} \cite{FKL}. It works in the specific models,
where mass of $\chi$ particles is proportional to inflaton field,
$m_{\chi}=g\phi$. When inflaton goes through minimum of potential $\phi=0$ 
$\chi$-particles are massless and they are very efficiently
produced. When $|\phi|$ increases, $m_{\chi}$ increases too and can
reach the value close to $m_{Pl}$. 

Another possible mechanisms of SHDM particle production are non-equilibrium 
thermal production at reheating and by early topological defects \cite{BKV}.
The latter can be produced at reheating \cite{KLS}.

{\em Gravitational production} of particles occurs due to time
variation of gravitational field during expansion of the universe 
\cite{ZeSt}. For particles with the conformal coupling with gravity,
$(1/6)RX$, where $R$ is the space-time curvature of the expanding 
universe, the particle mass itself couples a particle with the field 
(gravitation) and any other couplings are not needed. $X$ particles
can be even sterile! Neither inflation is needed for this production.
It rather limits the gravitational production of the particles. Since 
this production is described by time variation of the Hubble constant 
$H(t)$, only particles with masses $m_X \leq H(t)$ can be produced.
In inflationary scenario $H(t) \leq m_{\phi}$, where $m_{\phi}$ 
is the mass of the inflaton. It results in the limit on mass of 
produced particles $m_X \leq 10^{13}~GeV$ \cite{CKR,KuTk}. 
 
The gravitational production of superheavy particles was recently studied
in refs\cite{CKR,KuTk} (see \cite{KuTkr} for a review). It is
remarkable that for the mass 
$m_X \sim 10^{13}~GeV$ the relic density is $\Omega_X \sim 1 $  without
any additional assumptions. It makes superheavy particles most natural 
candidates for Cold DM. \\*[2mm]
\noindent
{\em Lifetime}\\
Superheavy particles are expected to be very short-lived. Even
gravitational interaction (e.g. described by dimension 5 operators suppressed
by the Planck mass) results in the lifetime much shorter than the age
of the Universe $t_0$. The superheavy particles must be protected
from fast decay by some symmetry, respected even by gravitational
interaction, and such symmetries are known. They are discrete gauge symmetries.
They can be very weakly broken e.g. by wormhole effects \cite{BKV} or
instanton effects \cite{KR} to provide the needed lifetime. The
systematic analysis of broken discrete gauge symmetries is given in 
ref.\cite{Ya}. For the group $Z_{10}$ the lifetime of X-particle with 
$m_X \sim 10^{13} - 10^{14}~GeV$ was found in the range 
$10^{11} - 10^{26}~yr$. The realistic elementary particle models for 
such long-lived particles were suggested \cite{BeEl,Ya1}.\\*[2mm]
\noindent
{\em Spectrum of UHECR}\\  
Quark and gluons produced in the decay of superheavy particle
originate QCD cascade, similar to that from $Z^0$ decay. The
resulting spectrum of hadrons can be calculated using the standard 
QCD technique \cite{Dok,Web}. The spectrum of hadrons is not
power-law, its most spectacular feature is the Gaussian peak at small
x. Photons dominate the primary spectrum by a factor $\sim 6$. 
The calculations of the spectrum were performed in ref.\cite{BiSa}
(HERWIG MC simulation for ordinary QCD) and in ref.\cite{BeKa}
(analytic MLLA calculations for SUSY QCD).\\*[2mm]
\noindent  
{\em Observational predictions}.\\
{\em Overdensity} $\delta$ of SHDM particles in the Galactic halo is the same 
as for any other form of CDM, and numerically it is given by a ratio
of CDM density  observed in the halo to CDM density in extragalactic 
space ($\delta \sim 10^5$). 

{\em Spectrum} of UHECR produced by decaying X-particles in the Galactic
halo and beyond is shown in Fig.2. One can see that UHE photon flux 
appreciably dominates over that of protons.
 
\vspace{-10mm}
\begin{figure}[htb]
\begin{center}
\psfig{bbllx=80pt, bblly=370pt, bburx=550pt, bbury=710pt,
file=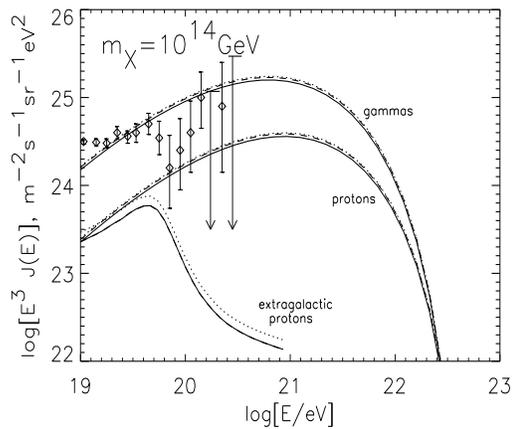, height=6cm, width=7cm, clip}
\end{center}
\vspace{-10mm}
\caption{\em Predicted fluxes of UHE photons and protons from the
decay of superheavy relic particles with mass $m_X=1\cdot 10^{14}~GeV$.
\cite{BBV}. The solid, dotted and dashed curves correspond to different
distributions of SHDM in the halo. Observational data are from AGASA.
} 
\end{figure}
\vspace{-7mm}
 
{\em Anisotropy} is caused by non-central position of the Sun in the
halo. Most notable effect, the difference in fluxes in directions 
of Galactic center and anticenter, cannot be observed by existing
arrays. Calculated phase and amplitude of the first harmonic of anisotropy 
\cite{BM,TW} are compared in Fig.3 with observations. In spite of the
visual agreement, one might only conclude that predicted anisotropy
does not contradict the observations: within $1.5\sigma$ AGASA data are
compatible with isotropy.
\newpage
\vspace{-10mm}
\begin{figure}[htb]
\begin{center}
\psfig{bbllx=130pt, bblly=430pt, bburx=440pt, bbury=730pt,
file=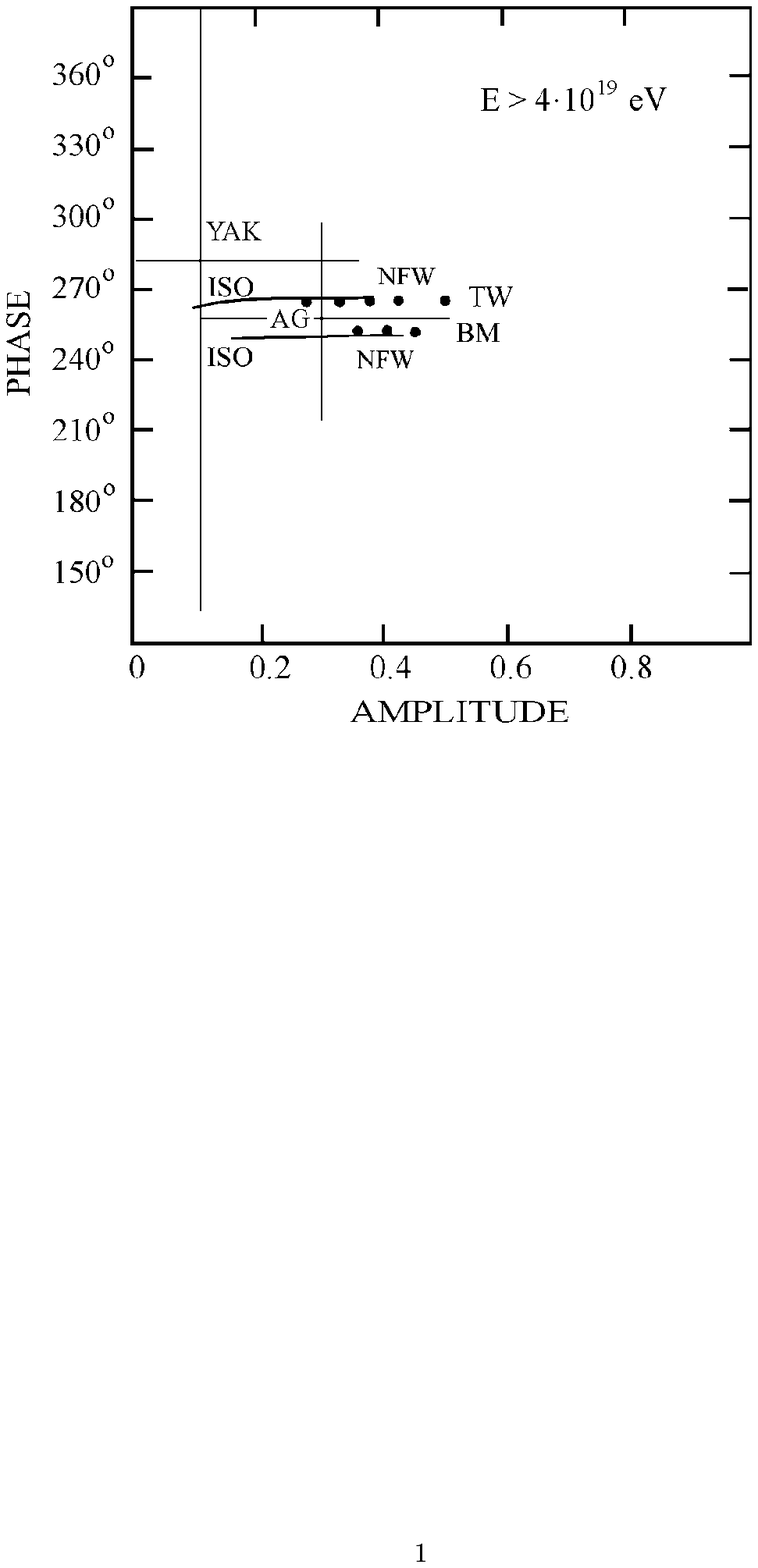, height=6cm, width=7cm, clip}
\end{center}
\vspace{-14mm}
\caption{\em Amplitude and phase of the first harmonic of anisotropy
for the AGASA (AG) and Yakutsk (YAK) arrays. Solid lines are for ISO 
distribution of DM and dots (NFW) for NFW numerical simulations \cite{NFW}. 
BM \cite{BM} and TW \cite{TW} calculations agree with each other.
} 
\end{figure}
\vspace{-3mm}
{\em Angular clustering} in UHECR arrival (doublet and triplet events) 
can be due to clumpiness of DM halo. Numerical N-body simulations show 
the presence of dense DM clouds in the halo. For example, the high
resolution simulations of ref.\cite{Moor} predict about 500 DM clouds 
with masses $M \sim 10^8 M_{\odot}$ in the halo of our Galaxy. The
baryonic content of these clouds should be low \cite{Kly}, and
therefore one cannot expect the identification of {\em all sources} of 
UHECR doublets and triplets with the observed clouds. The smallest
clumps resolved so far in the high resolution simulations reach 
$M_{cl} \sim 10^6 M_{\odot}$, {\em i.e.} they fall into range of the
globular cluster masses. It could be that some of the doublet/triplet 
UHECR sources are globular clusters. 

The high resolution simulations demonstrate the early origin of the
clumps ($ z \approx 5$) \cite{Moor}, and therefore the core overdensity, as
compared with present density, is $(1+z)^3 \sim 200$.  
The extended halos of DM clouds can be stripped away by tidal interactions
when clouds cross the galactic disc. The formation of 
dense compact DM objects was discussed in refs.\cite{Gur,Sil}.

Assuming the typical distance of a dense compact cloud to the sun as 
$r \sim 1~kpc$, one can estimate the fraction of UHE particles
arriving to us from one of these objects as 
$f \sim (M_{cl}/M_h)(R_h^2/r^2)$, where $R_h \sim 100~kpc$ is a size 
of the halo, and $M_{cl}$ and $M_h$ are the masses of a cluster and
halo, respectively. For $M_{cl} \sim 10^6 M_{\odot}, ~~ 
M_h \sim 10^{12} M_{\odot}$ and $r \sim 1~kpc$, one obtains $f \sim 0.01$,
{\em i.e.} about ten of such sources can provide the doublets and triplets
observed in AGASA and other detectors. Part of these sources can be
globular clusters. 

More detailed discussion will be presented in one of forthcoming
publications by A.Vilenkin and the author.

\section{Topological defects.}

{\em Topological defects, TD,} (for a review see \cite{Book}) can naturally 
produce particles of ultrahigh energies (UHE). The pioneering observation 
of this possibility was made by Hill, Schramm and Walker \cite{HS} (for 
a general analysis of TD as UHE CR sources see \cite {BHSS}).

In many cases TD become unstable and decompose to constituent fields, 
superheavy gauge and Higgs bosons (X-particles), which then decay 
producing UHECR. It could happen, for example, when two segments of 
ordinary string, or monopole and antimonopole touch each other, when 
electrical current in superconducting string reaches the critical value
and in some other cases.

In most cases the problem with UHECR from TD 
is not the maximal energy, but the fluxes. One very general reason 
for the low fluxes consists in the large distance between TD. A dimension
scale for this distance is the Hubble distance $H_0^{-1}$. However, in some 
rather exceptional cases this dimensional scale is multiplied to a small 
dimensionless value $r$. If a distance between TD is larger than 
 UHE proton attenuation length, then 
the flux at UHE is typically exponentially suppressed. 

The following TD have been discussed as potential sources of UHE
particles:
superconducting strings \cite{HS}, ordinary strings \cite{BR},
\cite{Vincent},\cite{WBMc}, magnetic monopoles , or more precisely 
bound monopole-antimonopole pairs (monopolonium \cite{Hill,BS}  and 
monopole-antimonopole connected by strings \cite{BO}),
networks of monopoles connected by strings \cite{BMV}, necklaces \cite{BV},
and vortons \cite{vorton}.

Monopolonia, monopole-antimonopole connected by strings and vortons
are clustering in Galactic halo \cite{BBV} and their observational
signatures for UHECR are identical to SHDM particles discussed above.

(i) {\em Superconducting strings}.\\
As was first noted by Witten\cite{Witten}, in a wide class of elementary 
particle models, strings behave like superconducting wires. Moving through 
cosmic magnetic fields, such strings develop electric currents.
Superconducting strings produce X particles when the electric current
in the strings reaches the critical value. Superconducting strings
produce too small flux of UHE particles \cite{BBV} and thus they are 
disfavoured as sources of observed UHECR.

(ii) {\em Ordinary strings}.\\
There are several mechanisms by which ordinary strings can produce UHE 
particles.

For a special choice of initial conditions, an ordinary  loop can collapse to a
double line, releasing its total energy in the form of X-particles\cite{BR}. 
However, the probability of this mode of collapse is
extremely small, and its contribution to the overall flux of UHE
particles is negligible.

String loops can also 
produce X-particles when they self-intersect (e.g. \cite{Shell}).
Each intersection, however, gives only a few
particles, and the corresponding flux is very small \cite{GK}. 

Superheavy particles with large Lorentz factors can be produced in 
the annihilation of cusps, when the two cusp segments overlap \cite{Bran}.  
The energy released in a single cusp event can be quite large, but
again, the resulting flux of UHE particles is too small to account for
the observations \cite {Bhat89,GK}.

It has been recently argued \cite{Vincent} that long
strings lose most of
their energy not by production of closed loops, as it is generally
believed, but by direct emission of heavy X-particles.
If correct, this claim will change dramatically 
the standard picture of string evolution. It has been also
suggested that the decay products of particles produced in this
way can explain the observed flux of UHECR \cite{Vincent}. 
However, as it is argued in ref \cite{BBV}, numerical simulations described in
\cite{Vincent} allow an alternative interpretation not connected with 
UHE particle production.  

But even if the conclusions of \cite{Vincent} were correct, the
particle production mechanism suggested in that paper cannot explain
the observed flux of UHE particles. If particles are emitted directly
from long strings, then the distance between UHE particle sources $D$ is
of the order of the Hubble distance $H_0^{-1}$, $D \sim H_0^{-1} \gg R_p$, 
where $R_p$ is the proton attenuation length  in the microwave background 
radiation. In this case UHECR flux has an exponential cutoff at energy 
$E \sim 3\cdot 10^{10}~GeV$. In the case of accidental proximity of a
string to the observer, the flux is strongly anisotropic. A fine-tuning 
in the position of the observer is needed to reconcile both 
requirements.

(iii) {\em Monopolonium and $M\bar{M}$-pair connected by string}.\\
Monopole-antimonopole pairs ($M\bar{M}$) can form bound state \cite{Hill}. 
Spiraling along the classical orbits  they fall to each other and 
annihilate, producing superheavy particles. The lifetime of this 
system depends on the initial (classical) radius, and can be larger
than the age of the Universe $t_0$ \cite{Hill}. Production of UHECR 
by monopolonia was studied in ref.\cite{BS} (clustering of monopolonia 
in the Galactic
halo was not noticed in this paper and was indicated later in ref.\cite{BBV}).

Recently \cite{BO} it was demonstrated that friction of monopoles in
the cosmic plasma results in the monopolonium lifetime much shorter
than $t_0$. Instead of monopolonium the authors have suggested  a
similar object, $M\bar{M}$ pair connected by a string, as a candidate
for UHECR. This TD is produced in the sequence of the symmetry breaking
$ G\to H\times U(1)\to H$. At the first symmetry breaking monopoles
are produced, at the second one each $M\bar{M}$-pair is connected by a
string. For the light strings the lifetime of this TD is larger than
$t_0$. $M\bar{M}$-pairs connected by strings are accumulated in the
halo as CDM and have the same observational signatures as SHDM particles.

(iv){\em Network of  monopoles connected by strings}.\\
The sequence of phase transitions
\begin{equation}
G\to H\times U(1)\to H\times Z_N
\label{eq:symm}
\end{equation}
 results in the formation of monopole-string networks in which each monopole 
is attached to N strings. Most of the monopoles and most of the strings belong 
to one infinite network. The evolution of networks is expected to be 
scale-invariant with a characteristic distance between monopoles 
$d=\kappa t$, where $t$ is the age of Universe and $\kappa=const$. 
The production of UHE particles are considered in \cite{BMV}. Each 
string attached 
to a monopole pulls it with a force equal to the string tension, $\mu \sim 
\eta_s^2$, where $\eta_s$ is the symmetry breaking vev of strings. Then
monopoles have a typical acceleration $a\sim \mu/m$, energy $E \sim \mu d$ 
and Lorentz factor $\Gamma_m \sim \mu d/m $, where $m$ is the mass of the 
monopole. Monopole moving with acceleration can, in principle, radiate  
gauge quanta, such as photons, gluons and weak gauge bosons, if the
mass of gauge quantum (or the virtuality $Q^2$ in the case of gluon) is
smaller than the monopole acceleration. The typical energy of radiated quanta 
in this case is $\epsilon \sim \Gamma_M a$. This energy can be much higher 
than what 
is observed in UHECR. However, the produced flux (see \cite{BBV}) is much 
smaller than the observed one. 

(v){\em Vortons}.\\
Vortons are charge and current carrying loops of superconducting
string stabilized by their angular momentum \cite{mash}.  Although
classically stable, vortons decay by gradually losing charge carriers
through quantum tunneling.  Their lifetime, however, can be greater
than the present age of the universe, in which case the escaping
$X$-particles will produce a flux of cosmic rays.  The $X$-particle
mass is set by the symmetry breaking scale $\eta_X$ of string 
superconductivity.  

The number density of vortons formed in the early universe is rather
uncertain.  According to the analysis in ref.\cite{BCDT}, vortons are
overproduced in models with $\eta_X > 10^9 GeV$, so all such models
have to be ruled out.  In that case, vortons cannot contribute to the
flux of UHECR.  However, an alternative analysis \cite{mash} suggests
that the excluded range is $10^9 GeV <\eta_X < 10^{12}GeV$, while for
$\eta_X \gg 10^{12}GeV$ vorton formation is strongly suppressed.  This
allows a window for potentially interesting vorton densities
with $\eta_X \sim 10^{12}-10^{13}GeV$. Production of Ultra High Energy 
particles by decaying vortons was studied in ref.\cite{vorton}.  

Like monopoles connected by strings and SH relic
particles, vortons are  clustering in the
Galactic halo and UHECR production and spectra are identical in these three 
cases. 

(vi){\em Necklaces}.\\
Necklaces are hybrid TD corresponding to the case $N=2$ in 
Eq.(\ref{eq:symm}), i.e. to the case when each monopole is attached to two
strings.  This system resembles ``ordinary'' cosmic strings,
except the strings look like necklaces with monopoles playing the role
of beads. The evolution of necklaces depends strongly on the parameter
\begin{equation}
r=m/\mu d,
\end{equation}
where $d$ is the average separation between monopoles and antimonopoles
along  the strings.
As it is argued in ref. \cite{BV}, necklaces might evolve to  
configurations with $r\gg 1$, though numerical simulations are needed to 
confirm this conclusion.  
Monopoles and antimonopoles trapped in the necklaces
inevitably  annihilate in the end, producing first the heavy  Higgs and 
gauge bosons ($X$-particles) and then hadrons.
The rate of $X$-particle production can be estimated as \cite{BV} 
\begin{equation}
\dot{n}_X \sim \frac{r^2\mu}{t^3m_X}.
\label{eq:xrate}
\end{equation}

Restriction due to e-m cascade radiation demands the cascade energy density 
$\omega_{cas} \leq 2\cdot 10^{-6}~eV/cm^3$. The cascade energy density 
produced by necklaces can be calculated as
\begin{equation}
\omega_{cas}=
\frac{1}{2}f_{\pi}r^2\mu \int_0 ^{t_0}\frac{dt}{t^3}
\frac{1}{(1+z)^4}=\frac{3}{4}f_{\pi}r^2\frac{\mu}{t_0^2},
\label{eq:n-cas}
\end{equation}
where $f_{\pi}\approx 0.5$ is a fraction of total energy release 
transferred to the cascade.
The separation between necklaces is given by \cite{BV} 
$D \sim r^{-1/2}t_0$ for large $r$. Since $r^2\mu$ is limited by cascade 
radiation, Eq.(\ref{eq:n-cas}), one can obtain a lower limit on the 
separation $D$ between necklaces as
\begin{equation}
D \sim \left( \frac{3f_{\pi}\mu}{4t_0^2\omega_{cas}}\right)^{1/4}t_0
>10(\mu/10^6~GeV^2)^{1/4}~kpc,
\label{eq:xi}
\end{equation}

Thus, necklaces can give a realistic example of the case when separation 
between sources is small and the Universe can be assumed  uniformly filled by 
the sources. 

The fluxes of UHE protons and photons are shown in Fig.4 according to 
calculations of ref.\cite{BBV}.
Due to absorption of UHE photons the 
proton-induced EAS from necklaces strongly dominate over those induced by 
photons at all 
energies except $E> 3\cdot 10^{11}~GeV$, where photon-induced 
showers can comprise an appreciable fraction of the total rate.

\vspace{-10mm}
\begin{figure}[htb]
\begin{center}
\psfig{bbllx=53pt, bblly=354pt, bburx=557pt, bbury=715pt,
file=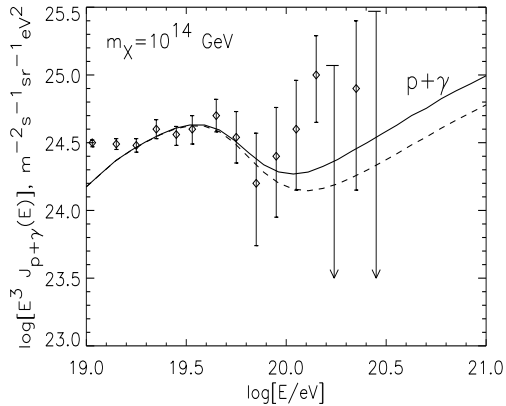, height=6cm,width=7cm}
\end{center}
\vspace{-14mm}
\caption{\em Spectrum of UHE photons + protons from necklaces
compared with AGASA measurements. Solid and dashed curves correspond
to different absorption of UHE photons in extragalactic space.
} 
\end{figure}
\vspace{-12mm}
\section{\bf Conclusions}

At $E\geq 1\cdot 10^{19}~eV$ a new component of cosmic rays with a flat 
spectrum is observed. 
According to the Fly's Eye and 
Yakutsk data the chemical composition is better described by protons than 
heavy nuclei. The AGASA data are consistent with isotropy in arrival 
of the particles, but about $20\%$ of particles at 
$E \geq 4\cdot 10^{19}~eV$ arrive as doublets and triplets within 
$\sim 2 - 4^{\circ}$.

The galactic origin of UHECR due to conventional sources is disfavoured: 
the maximal observed energies
are higher than that calculated for the galactic sources, and the
strong Galactic disc 
anisotropy is predicted even for the extreme magnetic fields in the disc 
and halo.

The signature of extragalactic UHECR is GZK cutoff. The position of 
steepening is model-dependent value. For the Universe uniformly filled with 
sources, the steepening starts at $E_{bb} \approx 3\cdot 10^{19}~eV$ and 
has $E_{1/2} \approx 6\cdot 10^{19}~eV$ (the energy at which spectrum 
becomes a factor of two lower than a power-law extrapolation from lower 
energies). The spectra of UHE nuclei exhibit steepening approximately at the 
same energy as protons. UHE photons have small absorption length due to 
interaction with radio background radiation. 

The extragalactic astrophysical sources theoretically studied so far, 
have either too small $E_{max}$ or are located too far away. The Local 
Supercluster (LS) model can give spectrum with $E_{1/2} \sim 10^{20}~eV$, if 
overdensity of the sources  is larger than 10. However, IRAS galaxy 
counts give overdensity $\delta=1.4$. 

GRBs and a nearby single source ({\em e.g.} M87) remain the potential
candidates for the observed UHECR.

Superheavy Dark Matter can be the source of observed UHECR. These
objects can be relic superheavy particles or topological defects
such as $M\bar{M}$-pairs connected by strings or vortons. These objects  
are accumulated in the halo and thus the resulting spectrum of UHECR does not
have the GZK cutoff. In this case UHECR is a signal from inflationary
epoch, because both superheavy particles and topological defects are
most probably produced during reheating.

The observational signatures of UHECR from SHDM are (i) absence of GZK
cutoff, (ii) UHE photons as the primaries and (iii) anisotropy due to 
non-central position of the Sun in the halo. The angular clustering is
possible due to clumpiness of DM in the halo.

Topological Defects naturally produce particles with extremely high 
energies, much in excess of what is presently observed. However, the fluxes 
from most known TD are too small. Only necklaces, $M\bar{M}$
connected by strings and vortons remain candidates for the sources of 
the observed UHECR. Necklaces give so far the only known example of
extragalactic TD as a sources of UHECR. Its signature is the 
presence of the photon component in the primary radiation and its dominance 
at the highest energies $E> 10^{20}~eV$.

\section{\bf Acknowledgments}

I am grateful to my co-authors Pasquale Blasi, Michael Kachelriess and 
Alex Vilenkin for many useful discussions.

\end{document}